\documentclass[preprint,aps,amsmath,superscriptaddress,tightenlines]{revtex4}
\usepackage{graphicx}
\usepackage{bm}
\usepackage{epsfig}

\def\Dslash{D\!\!\!\!\slash}

\def\ppslash{p^{\,\prime}\!\!\!\!\!\slash}
\def\nslash{n\!\!\!\slash}
\def\vslash{v\!\!\!\slash}
\def\bnslash{\bar n\!\!\!\slash}
\def\pslash{p\!\!\!\slash}
\def\kslash{k\!\!\!\slash}
\def\epsslash{\varepsilon\!\!\!\slash}

\def\vslash{v\!\!\!\slash}

\def\Aslash{A\!\!\!\slash}
\def\cAslash{{\cal {A\!\!\!\!\hspace{0.04cm}\slash}}}
\def\OMIT#1{}

\newcommand{\nn}{\nonumber}

\newcommand{\bn}{{\bar n}}
\newcommand{\bea}{\begin{eqnarray}}
\newcommand{\eea}{\end{eqnarray}}
\newcommand{\beq}{\begin{equation}}
\newcommand{\eeq}{\end{equation}}

\newcommand{\cPslash}{ {\cal P}\!\!\!\slash}

\newcommand{\mcdot}{\!\cdot\!}

\begin{document}

\preprint{ \hbox{ZU-TH 17/02} }

\title{\phantom{x}\vspace{0.5cm}
Factorization in leptonic radiative $B\to \gamma e\nu$ decays
\vspace{0.5cm} }

\author{Enrico Lunghi} \affiliation{Deutsches Elektronen Synchrotron,
  DESY, Notkestrasse 85, 22607, Hamburg, Germany\footnote{Present
    address: Institut f\"ur Theoretische Physik, Universit\"at
    Z\"urich, 8057 Z\"urich, Switzerland}}

\author{Dan Pirjol}
\affiliation{Department of Physics, University of California at San Diego,
        La Jolla, CA 92093\footnote{Present address: Department of Physics, 
Johns Hopkins University, 3400 N. Charles Street, Baltimore, MD 21218}}

\author{ Daniel Wyler}
\affiliation{Institut f\"ur Theoretische Physik, Universit\"at Z\"urich,
8057 Z\"urich, Switzerland}

\date{\today\\\vspace{1cm} }

\begin{abstract}
We discuss factorization in exclusive radiative leptonic $B\to\gamma 
e\bar\nu$ decays using the soft-collinear effective theory. The form factors
describing these decays can be expanded in a power series in 
$\Lambda_{QCD}/E_\gamma$ with $E_\gamma$ the photon energy.
We write down the most general
operators in the effective theory which contribute to the form factors
at leading order in $\Lambda_{QCD}/E_\gamma$, proving
their factorization 
into hard, jet and soft contributions, to all orders in $\alpha_s$. 
\end{abstract}

\maketitle
\newpage



\section{Introduction}

Recently, there has been substantial progress in the notoriously
difficult theoretical treatment of exclusive hadronic decays of
$B$-mesons. One one hand, it was shown that in the heavy quark limit
and in low orders of QCD 'factorization' holds in certain
processes~\cite{BBNS1,BBNS2}; on
the other hand, the soft-collinear effective theory (SCET) formulated
recently in Refs.~\cite{bfl,bfps,cbis,bpssoft} promises a new
systematic way to tackle all orders in the strong coupling. Such 
applications were discussed in \cite{bpsfact,ChayKim,BCDF,bpssoft,bfprs}.

Beside purely hadronic two-body decays such as $B \to K \pi$, also the
simpler exclusive radiative decays are of great interest.  For
instance, the rare radiative weak decays $b\to (s,d)\gamma$ are an
important source of information about the Standard Model and may
provide a window to possible 'New Physics'. For this reason their
study has received a great deal of attention, both experimentally and
theoretically.

While the theory of inclusive $b\to (s,d)\gamma$ decays is relatively
well under control, the corresponding exclusive decays are less
understood, although several model calculations of the matrix elements
exist (see, for instance, Refs.~\cite{sumrules,wa,aaw}) and some progress
towards the factorization of these matrix elements has been achieved
recently~\cite{befe,BeFeSe,BoBu,AliParkh} in the framework of the QCD
factorization for exclusive processes.

The simplest $B$ decay involving a hard photon is the radiative
leptonic decay $B\to\gamma e\nu$. This decay is not only a testing
ground for the various new methods of tackling exclusive decays, but
is also interesting in its own right because it is a clear probe of
some of the $B$ meson properties and weak
couplings~\cite{BuGoWy,CoFaNa,EHM,GLZ,KPY,HNLi}.
The motivation for the present study is connected with the fact that
many ingredients required for the hadronic decay $\bar B\to V\gamma$
already appear in this case but in a simplified setting.

The goal of the present paper is therefore to present a complete
all-order (in $\alpha_s$) proof of the factorization ansatz,
\bea
\label{WAfactor}
A (\bar B \to \gamma e \bar \nu) = C(E_\gamma/m_b) \cdot J \otimes \Psi_B
\eea
using the soft-collinear effective theory formulated
recently in Refs.~\cite{bfl,bfps,cbis,bpssoft}. Here, $C$
stands for the Wilson coefficients of the SCET operators and encode
the hard gluons effects, $J$ are the so-called jet functions and take
into account contributions from collinear loop momenta, $\Psi_B$ is the
$B$-meson wave function and includes ultra-soft gluons effects;
$\otimes$ denotes a convolution along the light-cone component of the
spectator momentum ($k_+= n\mcdot k$). This decay was first considered in
QCD factorization in \cite{KPY}, and more recently in Ref.~\cite{DGS}, 
where a factorization formula of type (\ref{WAfactor}) is proved at 
one-loop order.

The effective theory greatly simplifies factorization proofs of
perturbative QCD, by reducing a complicated diagrammatic analysis to
simple field transformations on operators in the effective
Lagrangian~\cite{bpssoft,bfprs}.  Using SCET methods, we first identify the
operators contributing to this decay and then prove the factorization
of the hadronic matrix elements for this process in a form similar to
(\ref{WAfactor}) to all orders in $\alpha_s$.

The contents are organized as follows. The effective theory is briefly
summarized in Sec.~II, where we recapitulate the basic notions needed
for the rest of the paper. In Sec.~III, we consider the leptonic
radiative $B$ decay $\bar B\to \gamma e\nu$. We show, in particular,
that the form factors describing this decay can be written as a
product of hard, collinear and soft factors. The symmetries of the
effective theory give symmetry relations among $B\to\gamma$ form factors, 
to all orders in $\alpha_s$. We find that the five form factors describing
these decays reduce at leading order in $\Lambda/E_\gamma$ to just one
independent function, proving a result conjectured in \cite{KPY} from a
one-loop computation. Necessary technical details are presented in the
Appendix.

\section{Basics of the soft-collinear effective theory (SCET)}
The effective theory developed in \cite{bfl,bfps,cbis,bpssoft}
describes particles (quarks and gluons) relevant for decays of a heavy
meson into fast light hadrons. The relevant scales in this problem are
$Q, Q\lambda$ and $Q\lambda^2$, where $Q$ is the large energy of the
final fast hadron $\simeq m_b$ and $\lambda^2 \simeq p_\perp^2/Q^2$
with $p_\perp$ the typical transverse momentum of the particles. The
goal is to obtain an expansion in powers of $\lambda$; to this extent,
the $\lambda$-scaling of momenta and fields is important.

Referring to the papers \cite{bfl,bfps,cbis,bpssoft} for a detailed
explanation of the theory, we present here the few basic points that
will be required in the following. We take the hadrons moving with
large momentum and small invariant mass in a light-cone direction,
$n$. The other light-cone vector, $\bar n$, is chosen such that $n^2 =
{\bar n}^2 = 0$ and $n \mcdot \bar n = 2$. Momenta are decomposed along the
light cone as $p=(n \mcdot p, \bar n \mcdot p, p_\perp) = (p^+, p^-, p_\perp)$
with
\bea
p^\mu = \frac{n\mcdot p}{2} \bar n^\mu + \frac{\bar n \mcdot p}{2} n^\mu + 
p_\perp^\mu,
\eea
In the effective theory one introduces distinct fields for each
relevant momentum region. These include the following: collinear modes
for quarks ($\xi_{n,p}$) and gluons ($A^\mu_{n,q}$) with momenta
scaling as $Q(\lambda^2,1,\lambda)$; ultra-soft modes ($A_{\rm us}$
and $q_{\rm us}$) with momenta scaling as $Q(\lambda^2, \lambda^2,\lambda^2)$
and soft modes whose momenta scale as $Q(\lambda, \lambda,\lambda)$.
The fields also have a well definite scaling in powers of $\lambda$:
$A_{n,q} \sim (\lambda^2, 1, \lambda)$, $A_{\rm us}\sim \lambda^2$,
$\xi_{n,p} \sim \lambda$ and $q_{\rm us} \sim \lambda^3$. 

The collinear fields are related to the full theory fields 
by \footnote{See Ref.~\cite{BCDF} for a somewhat different treatment of the
collinear fields.}
\bea
\phi(x) = \sum_{\tilde p} e^{-i\tilde p x}\phi_{n,p} (x)
\eea
where $\tilde p = \frac{1}{2}(\bar n \mcdot p) n + p_\perp$ is the
large part of the momentum, treated as a label on the collinear field. 
Although in the following we will
for convenience often write $p$ instead of $\tilde p$ in the exponents,
one should always keep in mind that algebraic manipulations of
exponents (and fields) should only involve the appropriate components.

It is convenient to introduce a `label' operator ${\cal P}$ \cite{cbis}
which acts on the collinear fields and picks up their large momentum
${\cal P}^\mu\xi_{n,p} = (\frac{\bar n \mcdot p}{2} n^\mu + p_\perp^\mu)
\xi_{n,p}$. We will frequently use a special notation which associates
a well-defined momentum label index to an arbitrary product of
collinear fields. This is defined according to
\bea
\label{label}
[\phi^{(1)}_{n}\cdots \phi^{(j)}_{n}]_p \equiv \delta_{\bn\mcdot p,
\bn\mcdot {\cal P}} \; [\phi^{(1)}_{n}\cdots \phi_{n}^{(j)}]
\eea
where $\delta_{\bn\mcdot p, \bn\mcdot {\cal P}}$ acts only inside the
square brackets. In this and in the following expressions, we usually
omit the momentum labels of the collinear fields $A_{n,q}$ and $\xi_{n,p}$.
Moreover, appropriate summation over the (omitted) labels of the
fields is usually implied. 

The longitudinal component of collinear gluons moving in the $n_\mu$
direction, $\bn\mcdot A_{n,q}$, appears only in the combination $W_n =
\exp(-\frac{g}{\bn\mcdot{\cal P}} \bn \mcdot A_{n})$ which is
essentially a Wilson line along the $\bn$ light cone direction. The
most general gauge invariant collinear operators can be built using
the $\xi_{n}$, $A_n$, $W_n$ and ${\cal P}_\mu$ building blocks.

Particular products of collinear fields appear often; for convenience
of writing we will sometimes use
\beq\label{special}
\chi_{n} \equiv W_n^\dagger \xi_{n}\,,\qquad
\Phi_{n} \equiv W_n^\dagger (\cPslash_\perp + g\Aslash_{n\perp})
\frac{\bnslash}{2}\xi_{n}.
\eeq

The precise definition of the expansion parameter $\lambda$ is
specific to each problem. For the $\bar B\to \gamma e\bar\nu$ decay
discussed in Sec.~III, $\lambda$ is defined by the typical virtuality
of the light quark struck by the emitted photon, $p^2 \sim
Q^2\lambda^2 \sim E_\gamma\Lambda_{QCD}$ with $E_\gamma$ the photon
energy in the rest frame of the $B$ meson, which gives $\lambda^2 \sim
\Lambda_{QCD}/E_\gamma$.  With this definition, the typical momentum
of the spectator in the $B$ meson is of order $\Lambda_{QCD} \sim
Q\lambda^2$, such that its momentum is ultra-soft. 
On the other hand, for the $\bar B\to D\pi$ decay, the typical virtuality 
of the collinear partons in the pion is of order $\Lambda^2_{QCD}$, leading 
to $\lambda=\Lambda_{QCD}/m_b$; in this case, the momenta of the $B$ 
constituents are soft \cite{bpsfact,bpssoft}.

The interactions  among the effective theory fields
are described by an effective Lagrangian. It consists basically
of the most general combination of
fields and/or suitable products of fields (such as in Eq.~(\ref{special})),
compatible with gauge invariance and reparameterization invariance
\cite{ChayKim,mmps}. This Lagrangian
is organized in increasing powers of $\lambda$.
Similarly, the currents of the full theory are matched onto operators in the 
effective theory, and can be obtained by matching to the full theory, order by
order in $\lambda$.

The couplings of the collinear quark field $\xi_{n,p}$ to ultra-soft and
collinear gluons $A_{us}$ and $A_{n,p}$, respectively, are described by
the soft-collinear effective Lagrangian.  At leading order $O(\lambda^0)$, this is
given by \cite{bfps}
\bea\label{LSCET}
{\cal L}^{(0)} = \bar \xi_n \left\{
n\mcdot iD + g\;n\mcdot A_n + (\cPslash_\perp + 
g\; \Aslash_{n\perp})\frac{1}{\bn\mcdot {\cal P}
+ g\;\bn\mcdot A_n}
(\cPslash_\perp + g\;\Aslash_{n\perp})\right\}\frac{\bnslash}{2}\xi_n\,.
\eea
Here, the covariant derivative $D_\mu = \partial_\mu - ig A_\mu^a T^a$ 
contains ultra-soft gauge fields $A_\mu$ only.

The form of operators in the effective theory is severely restricted
by collinear gauge invariance. These constraints have been discussed
at length in~\cite{cbis,bpssoft} for the case of operators
contributing to semileptonic and non-leptonic weak decays.  For
processes containing one hard photon we will find it useful to
consider invariance under the combined $SU(3)$ and $U(1)_{\rm e.m.}$
collinear gauge transformations. Although we work only to first order
in the electromagnetic coupling, this provides a convenient way of
automatically incorporating electromagnetic couplings in a gauge
invariant way.

To this extent, we consider the couplings of a charged collinear quark
moving along the $n$ direction $\xi_n$ to collinear gluons $A_n$ and
photons ${\cal A}_n$ moving along the same direction. A simple
extension of the argument presented in \cite{cbis,bpssoft} shows that
the latter must appear at lowest order in the combination
\bea W[\bn\mcdot A_n,
\bn\mcdot {\cal A}_n] = \exp\left[\frac{1}{\bn\mcdot {\cal P}}
(-g\;\bn\mcdot A_n - e\; Q\; \bn\mcdot {\cal A}_n)\right] \; ,
\eea
where $Q$ is the electric charge matrix defined in flavor space.
Under a general e.m. $n$-collinear gauge transformation
\beq
{U} = \sum_P e^{-iP\mcdot x} {U}_P(x) \;,
\eeq
with ${U}_P$ a matrix in flavor space, the charged $n$ collinear
fields transform according to
\bea
\xi^{(q)}_{n,p} &\to& {U}^{(q)}_{p-P}\xi_{n,P}^{(q)}\,,\qquad
W \to {U}_P W \; .
\eea
Finally, the heavy quarks are treated according to the heavy quark effective
theory. We write the generic heavy quark field with
four-velocity $v$ as $Q(x) = \exp{(-i m v \mcdot x)}\,Q_v(x)$ and
decompose it into a large and small component $Q_v = h_v + H_v$, with
\bea
h_v = \frac{1 + \vslash }{2}Q_v \,.
\eea

\section{Factorization in $B\to \gamma e\nu$ decays}

The simplest $B$ decay involving a hard photon is the radiative
leptonic decay $\bar B\to\gamma e \bar\nu$. 
Our main result will be to show that the $\bar B\to \gamma e \bar \nu$
amplitude factorize according to
\beq
A(\bar B\to \gamma e \bar\nu) =  C(E_\gamma/m_b) \int \frac{\mbox{d} n\mcdot k}{4\pi}
J(n\mcdot k) \Psi_B(n\mcdot k)
\eeq
where $C(E_\gamma/m_b)$ is an effective theory Wilson coefficient, $J$ is a
collinear function and $\Psi_B$ is the usual light-cone $B$ meson wave
function. Using the effective theory formulation we will prove also 
symmetry relations between form factors parametrizing $B\to \gamma$
transitions.

The $B\to\gamma e\bar\nu$ amplitude is given in QCD by
\bea\label{A}
A( B^-(v)\to\gamma(q,\varepsilon) e\bar\nu) &=& \nn \\
  & & \hskip - 3.5cm ie Q_q \frac{4G_F}{\sqrt2} V_{ub}\int d^4 x e^{iq\mcdot x}
\langle 0|\mbox{T}([\bar q\gamma_\mu P_L b](0),
[\bar q \epsslash^* q](x))|\bar B(v)\rangle\,
[\bar u(p_e)\gamma_\mu P_L v(p_\nu)]
\eea
with $P_L = \frac12 (1-\gamma_5)$.

We fix the kinematics by taking the photon momentum as $q_\mu =
E_\gamma n_\mu$ with the light-cone vector given by $n = (1,0,0,1)$.
The direction orthogonal to $n$ along the light-cone is parameterized
by $\bn = (1,0,0,-1)$. The electromagnetic current $\bar q \epsslash^*_\perp
q$ takes the ultra-soft spectator anti-quark in the $B$ meson carrying
momentum $k\simeq \Lambda_{QCD}$ into a collinear quark with a large
momentum component along the $n$ light-cone direction $q-k$.

The weak current $\bar q\Gamma b$, coupling a heavy quark to an
energetic light quark, is matched in SCET according to~\cite{bfps}
\bea\label{Jheavy}
\bar q\Gamma b = \sum_i C_i(\bn\mcdot p, \mu)
\bar \chi_{n,p} \Gamma_i h_v + O(\lambda)
\eea
with $C_i(\omega, \mu)$ Wilson coefficients. The explicit choice of
the Dirac structure in these operators differs from that used in \cite{bfps}, 
but they are simply related by algebraic manipulations.
To leading order in $\lambda$, the
vector and axial currents are matched onto the effective theory
operators (with $\omega = \bn\mcdot p$)
\bea
\bar q\gamma_\mu b &=& C_1^{(v)}(\omega,\mu) \bar\chi_{n,\omega}\gamma_\mu h_v +
C_2^{(v)}(\omega,\mu) \bar\chi_{n,\omega}v_\mu h_v +
C_3^{(v)}(\omega,\mu) \bar\chi_{n,\omega}n_\mu h_v\\
\bar q\gamma_\mu\gamma_5 b &=&
C_1^{(a)}(\omega,\mu) \bar\chi_{n,\omega}\gamma_\mu\gamma_5 h_v +
C_2^{(a)}(\omega,\mu) \bar\chi_{n,\omega}v_\mu\gamma_5 h_v +
C_3^{(a)}(\omega,\mu) \bar\chi_{n,\omega}n_\mu\gamma_5 h_v\,.
\eea
Explicit results for the Wilson coefficients $C_i^{(v,a)}(\omega,\mu=m_b)$
at one-loop order can be found in
Eq.~(33) of Ref.~\cite{bfps}. Only two of these Wilson coefficients are needed
in the following
\bea
C_1^{(v)}(\omega,m_b) &=& C_1^{(a)}(\omega,m_b) = 
1 - \frac{\alpha_s C_F}{4\pi} \left\{
2\log^2\left(\frac{\omega}{m_b}\right) + 2\mbox{Li}_2(1-\frac{\omega}{m_b}) 
\right.\\
& &\left.
+ \log\left(\frac{\omega}{m_b}\right)
\frac{3\omega-2m_b}{m_b-\omega} + \frac{\pi^2}{12} + 6
\right\}\nonumber
\eea

For completeness we give also the matching of the tensor current 
$\bar q\sigma_{\mu\nu} b$, which contributes to the amplitude for the
rare decay $B_s\to e^+e^- \gamma$ \cite{Bse+e-}. This is given at leading
order in $\lambda$ in terms of four Wilson coefficients $C_i^{(t)}(\omega,\mu)$
defined as
\bea
\bar qi\sigma_{\mu\nu} b &=& C_1^{(t)}(\omega,\mu) \bar\chi_{n,\omega}i\sigma_{\mu\nu} h_v +
C_2^{(t)}(\omega,\mu) \bar\chi_{n,\omega}(n_\mu\gamma_\nu - n_\nu \gamma_\mu) h_v\\
&+&
C_3^{(t)}(\omega,\mu) \bar\chi_{n,\omega}(v_\mu\gamma_\nu - v_\nu \gamma_\mu) h_v
+ C_4^{(t)}(\omega,\mu) \bar\chi_{n,\omega}(n_\mu v_\nu - n_\nu v_\mu) h_v \nonumber
\eea
The corresponding matching conditions at one-loop order are \cite{bfps}
\bea
C_1^{(t)}(\omega,m_b) &=& 
1 - \frac{\alpha_s C_F}{4\pi} \left\{
2\log^2\left(\frac{\omega}{m_b}\right) + 2\mbox{Li}_2(1-\frac{\omega}{m_b}) 
\right.\\
& &\left.
+ 2\log\left(\frac{\omega}{m_b}\right)
\frac{2\omega-m_b}{m_b-\omega} + \frac{\pi^2}{12} + 6
\right\}\nonumber\\
C_2^{(t)}(\omega,m_b) &=& \frac{\alpha_s C_F}{4\pi}
\frac{2\omega}{m_b-\omega} \log\left(\frac{\omega}{m_b}\right)\nonumber\\
C_3^{(t)}(\omega,m_b) &=& C_4^{(t)}(\omega,m_b) = 0\,.\nonumber
\eea

The matching of the light quark photon vertex $\bar q\epsslash_\perp^* q$
onto SCET operators is slightly more complicated. In addition to a
local term as in the case of the heavy-light current, this matching
contains also a non-local term:
\bea
\label{Jlight}
(\bar q \epsslash^*_\perp q)(x) = \varepsilon^\mu {\cal J}_\mu
 = \sum_p e^{-i p \mcdot x}
(\bar q \epsslash^*_\perp \chi_{n, p})(x) + i\int\mbox{d}^4 y
\mbox{T} ( {\cal J}_g(x)\,, {\cal L}_{q\xi}^{(1)}(y)).
\eea
The local term describes contributions in which the energetic photon
itself turns the ultra-soft quark $q$ directly into a collinear quark
$\xi_n$. The nonlocal term, on the other hand, describes the process
in which the $O(\lambda)$ subleading Lagrangian 
${\cal L}^{(1)}_{\xi q}$ merely turns the ultra-soft quark into a collinear 
quark; its coupling to the photon is then described by a new current ${\cal J}_g$.
The right hand side must be evaluated in the presence of the
interaction (\ref{LSCET}).

The expression for the subleading Lagrangian
${\cal L}_{q\xi}^{(1)}$ has been obtained in Ref.~\cite{BCDF} and
reads
\beq
{\cal L}_{q\xi}^{(1)}(y) = [\bar q
W^\dagger_n (\cPslash_\perp^\dagger + g\Aslash_{n\perp}) \xi_{n}](y) +h.c.
\eeq
The explicit form of ${\cal J}_g$ can be obtained by coupling the photon as a
collinear field and expanding the Lagrangian (\ref{LSCET}) to linear
order in $e$ (with $\Phi_{n,p}$ as defined in (\ref{special}))
\bea
\label{J0}
{\cal J}_g(x) = \sum_{p_1, p_2} e^{i(q-p_1+p_2)\mcdot x}
\left\{
[\bar\chi_{n,p_2}\epsslash_\perp^* \frac{1}{\bn\mcdot\cal P} \Phi_{n,p_1}] (x)
+ [\bar\Phi_{n,p_2}
\frac{1}{\bn\mcdot {\cal P} - \bn\mcdot q}
\epsslash_\perp^* \chi_{n,p_1}] (x)\right\} \, .
\eea

Note that, despite the $O(\lambda)$ suppression of 
${\cal L}_{q\xi}^{(1)}$, both terms in (\ref{Jlight}) contribute to the
same order in $\lambda$. The first term scales like $\lambda^4$ by simple
power counting of the fields \cite{bpspc}. In the
second term, ${\cal J}_g(x)$ scales like $\lambda^3$ and ${\cal L}_{q\xi}^{(1)}$
like $\lambda^5$. Taking into account a factor $\lambda^{-4}$ from
the $y$-integration yields a total scaling of $\lambda^4$ for the second term as well.
We have not added any new Wilson coefficients in
(\ref{Jlight}) since the form of the operators is restricted to all
orders by the arguments in \cite{BCDF}. 
This is also checked by direct computation in
the Appendix, where we show explicitly that, at one-loop order,
Eq.~(\ref{Jlight}) reproduces correctly the IR behavior of the full
QCD. 

Inserting Eqs.~(\ref{Jlight}) and (\ref{Jheavy}) into the time-ordered
product that appears in Eq.~(\ref{A}) one obtains the hard-collinear
factorization of the matrix element
\bea\label{hard}
A(\bar B(v)\to\gamma(q,\varepsilon) e\bar\nu) = ie Q_q \frac{4G_F}{\sqrt2} V_{ub}
\sum_{i=1}^6
\sum_{\bn\mcdot p} C_i(\bn\mcdot p)
A_{\rm eff}^i [\bar u(p_e) \gamma_\mu P_L v(p_\nu)]
\eea
where the decay amplitudes in the effective theory are given by
time-ordered products of ultra-soft and collinear fields
\bea\label{Aeff}
A_{\rm eff}^i &=&
\int d^4 x e^{i(q-p_2)\mcdot x}
\langle 0|\mbox{T} (
[\bar\chi_{n,p_1} \Gamma_i h_v](0)\,,
[\bar q \epsslash^*_\perp  \chi_{n, p_2}](x) )| \bar B(v)\rangle\\
&+&
i\int d^4 x d^4 y
\langle 0|\mbox{T} (
[\bar\chi_{n,p_1} \Gamma_i h_v](0)\,,
 {\cal J}_g(x)\,,
 {\cal L}_{q\xi}^{(1)}(y))
| \bar B(v)\rangle\,.\nonumber
\eea

In the next step the ultra-soft fields are decoupled from the collinear
fields $\xi_{n,p}$ and $A_n$. This is achieved via the field
redefinitions~\cite{bpssoft}
\bea\label{redef}
\xi_{n,p}(x) \to Y_n (x)\xi_{n,p}^{(0)}(x)\,,\qquad
A_n(x) \to Y_n (x) A_n^{(0)}(x) Y_n^\dagger(x)
\eea
where $Y_n$ is a Wilson line along the light cone direction $n$
containing the ultra-soft gluon field
\bea
Y_n(x) = \mbox{P}\exp\left( ig\int_{-\infty}^{x_-/2}
\mbox{d}\lambda n\mcdot A_{\rm us}(\lambda n)\right)\,.
\eea
The collinear fields with a $(0)$ superscript do not couple to
ultra-soft gluons anymore, and only interact with other collinear
fields.  Expressed in terms of the $\chi_{n,p}$ field, the
transformation (\ref{redef}) reads simply $\chi_{n,p}(x) \to Y_n(x)
\chi^{(0)}_{n,p}(x)$.  Similarly, the fields $\Phi_{n,p}$ introduced
in Eq.~(\ref{special}), transform as $\Phi_{n,p} \to Y_n
\Phi_{n,p}^{(0)}$. This simple transformation lies at the heart of
the factorization.
After applying the field redefinition (\ref{redef}) to
Eq.~(\ref{Aeff}), the time-ordered product can be factored into the
product of a collinear and an ultra-soft matrix element.

\begin{figure}[t!]
\begin{center}
 \includegraphics[height=3.5cm]{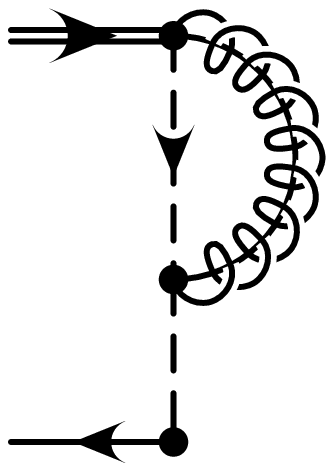}\hspace{1cm}
 \includegraphics[height=3.5cm]{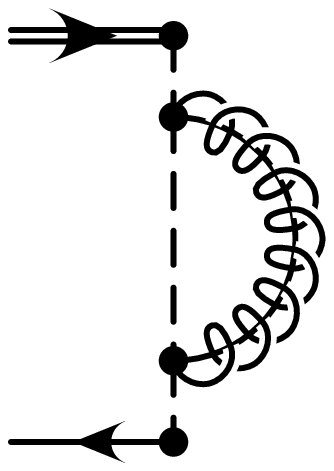}\hspace{1cm}
 \includegraphics[height=3.5cm]{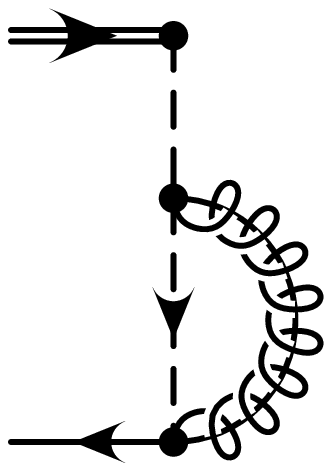}\hspace{1cm}
 \includegraphics[height=3.5cm]{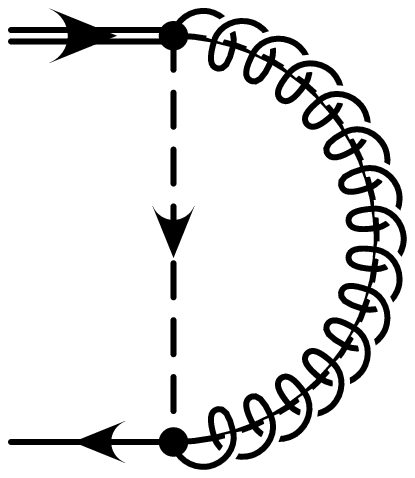}\\
(a) \hspace{3cm} (b) \hspace{3cm} (c) \hspace{3cm} (d) \\
\vspace{0.5cm}
  \includegraphics[height=3.5cm]{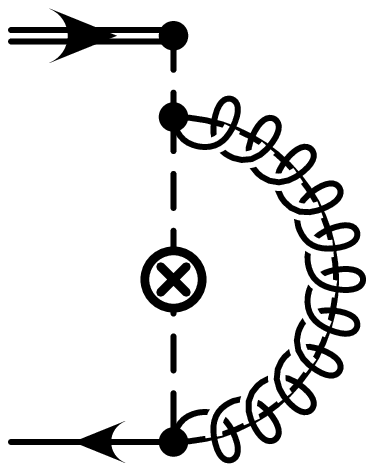}\hspace{1cm}
 \includegraphics[height=3.5cm]{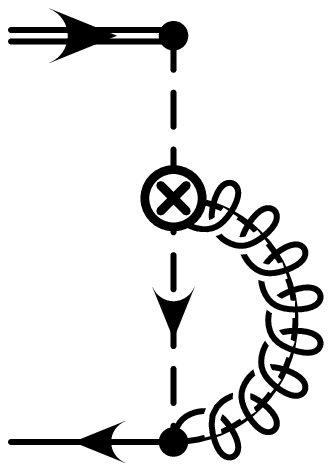}\hspace{1cm}
 \includegraphics[height=3.5cm]{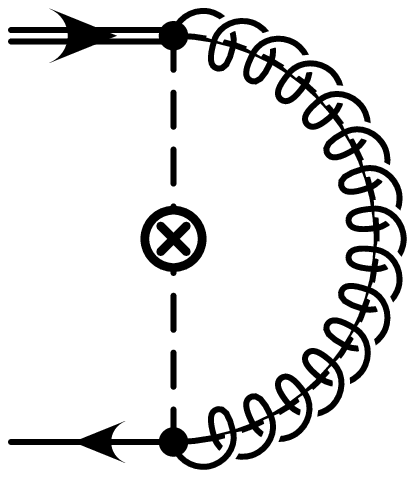}\\
 (e) \hspace{3cm} (f) \hspace{3cm} (g) 
\end{center}
{\caption{One-loop effective theory graphs contributing to the T-product 
(\ref{Aeff}). Only the collinear graphs are shown; 
the topology of the usoft graphs is identical to that of the QCD graphs. 
The graphs (a)-(d) are
produced by the first term in (\ref{Aeff}), and (e)-(g) by the second
term.
The lower vertex in (a)-(d) denotes the $\bar\xi \epsslash_\perp^* Wq$ operator;
in (e)-(g) the photon attaches to the circled cross vertex, denoting 
the  ${\cal J}_g$ operator, and the lower vertex is the
${\cal L}^{(1)}_{\xi q}$ subleading Lagrangian.
\label{collTprod}}}
\end{figure}

We start by considering the first term in Eq.~(\ref{Aeff}) which is
written explicitly as
\bea\label{Aeff2}
 & & \hskip -0.5 cm A_{\rm eff}^{i(1)} = \nn \\
 & & \hskip -0.5 cm  -\int \mbox{d}^4 x e^{i(q-p_2)\mcdot x}
\mbox{Tr}\, \left[
\langle 0|\mbox{T} (\chi_{n,p_2}^{(0)}(x), \bar \chi^{(0)}_{n, p_1}(0))|0\rangle
\Gamma_i \langle 0| \mbox{T} ([Y_n^\dagger h_v](0), [\bar q Y_n](x))
|\bar B(v)\rangle
\epsslash^*_\perp \right]\,.
\eea
The integration over $x$ allows to express the label momentum
$\bn\mcdot p_2$ in terms of the photon momentum as $\bn\mcdot p_2 =
\bn\mcdot q = 2E_\gamma$.

The collinear factor is the propagator of the $\chi_n$ fields, which,
in momentum space, reads~\cite{bpssoft}
\bea\label{Jdef}
& &\langle 0|\mbox{T}
(\chi_{n,p_2}^{\alpha (0)}(x), \bar \chi^{\beta(0)}_{n, p_1}(0))|0\rangle =
i\delta_{\bn\mcdot p_1, \bn\mcdot p_2}
\int \frac{d^4 k}{(2\pi)^4} e^{-ik\mcdot x}
J(\bn\mcdot p_1; n\mcdot k) \left(\frac{\nslash}{2}\right)_{\alpha\beta} \; .
\eea
This function was introduced in Ref.~\cite{bpssoft} in connection with
the factorization of the photon energy spectrum in the endpoint of
$B\to X_s\gamma$ decay. The discontinuity of $J(\bn\mcdot p; n\mcdot k)$ across its
cut gives the jet function appearing in the factorization formula for
$B\to X_s\gamma$. Here we need the function $J(\bn\mcdot p; n \mcdot k)$ itself in a
region away from the cut, where it can be computed in perturbation
theory provided that the virtuality $p^2 = \bn\mcdot p \; n \mcdot k$ is
sufficiently large, {\it i.e.} $p^2 \gg \Lambda_{QCD}^2$. The explicit
result at one loop order is~\cite{bfl,bfps} (with $L = \log(\bn\mcdot p
\; n\mcdot k/\mu^2)$)
\bea
J(\bn\mcdot p; n\mcdot k) = \frac{1}{n\mcdot k}\left[
1 + \frac{\alpha_s C_F}{4\pi}\left( 2L^2 - 3L + 7 - \frac{\pi^2}{3}.
\right)\right] \, .
\label{jetfunctionJ}
\eea
Note that the collinear function $J(\bn\mcdot p; n\mcdot k)$ depends on
$k$ only through $n\mcdot k$ to all orders in perturbation theory since
the soft--collinear Lagrangian contains only the $n\mcdot \partial$
derivative~\cite{bpssoft}: this implies that the coordinate $x$ in
Eq.~(\ref{Aeff2}) is constrained to be on the light cone ({\it i.e.}
$x_\perp = n \mcdot x = 0$). We obtain, finally, the factorized form of
the first term in Eq.~(\ref{Aeff}) as a one-dimensional integral
\bea\label{Aeff3}
A_{\rm eff}^{i(1)} = i\delta_{\bn\mcdot q, \bn\mcdot p_1}
\delta_{\bn\mcdot p_1, \bn\mcdot p_2}\int\mbox{d} \bn\mcdot x
\int \frac{\mbox{d}n\mcdot k}{4\pi} 
e^{\frac{i}{2}n\mcdot k \bn\mcdot x} J(\bn\mcdot p; n\mcdot k)
\mbox{Tr}\,
\left[\frac{\nslash}{2}
\Gamma_i \Psi_B(\bn\mcdot x)\epsslash^*_\perp \right]\,.
\eea
The ultra-soft matrix element on a $B$ meson defines its light-cone wave function
\bea
\Psi_B^{\alpha\beta}(\bn\mcdot x) 
&\equiv&
\langle 0| {\rm } T (\bar q^\beta({\bn\mcdot x\over 2} n), 
                     Y_n({\bn\mcdot x\over 2} n,0) , 
                     h_v^\alpha(0)
                    )|\bar B(v)\rangle \nonumber \\ 
&=&
\int \frac{\mbox{d}n\mcdot k}{4\pi} e^{-\frac{i}{2}n\mcdot k \bn\mcdot x} 
\Psi_B^{\alpha\beta}(n\mcdot k)
\eea
with $Y_n(x,0)=Y^\dagger_n(0) Y_n(x)$.

In the heavy quark limit, the
most general parameterization for $\Psi_B$ involves two functions
$\psi_\pm(n\mcdot k)$, which are usually defined as~\cite{BBNS1,BBNS2,grne,befe}
\bea\label{Psidef}
\Psi_B(n\mcdot k) =
\frac{1+\vslash}{2}\left( \frac{\nslash \bnslash}{4} \psi_+(n\mcdot k) +
\frac{\bnslash \nslash}{4} \psi_-(n\mcdot k)\right) \gamma_5\,.
\eea
With this definition, the light cone wave functions $\psi_\pm(n\mcdot
k)$ are normalized as
\bea
\int\frac{dn\mcdot k}{2\pi} \psi_\pm (n\mcdot k) = f_B m_B\,.
\eea
Expressed in terms of the momentum space wave functions, the first
term in the effective theory amplitude (\ref{Aeff}) is given in terms
of one of the two functions $\psi_\pm(n\mcdot k)$ and is written as
\bea\label{Aeff4}
A_{\rm eff}^{i(1)}(\bar B\to \gamma e\bar\nu) =
i\delta_{\bn\mcdot q, \bn\mcdot p_1}\delta_{\bn\mcdot p_1, \bn\mcdot p_2}
\int \frac{dn\mcdot k}{4\pi}
J(2E_\gamma; n\mcdot k) \psi_+(n\mcdot k) \mbox{Tr}\,
\left[\frac{\nslash}{2}
\Gamma_i \frac{1+\vslash}{2} \gamma_5 \epsslash^*_{\perp} \right].
\eea

The factorization of the second term in Eq.~(\ref{Aeff}) can be proved
in a similar way. After the field redefinition (\ref{redef}) the
relevant collinear matrix element is (the collinear field $\Phi_{n,p}$
is defined in Eq.~(\ref{special}))
\bea\label{Jell}
& &i\int \mbox{d}^4 x e^{i(q-p_1+p_2)\mcdot x}
\langle 0|
\mbox{T}(\Phi_{n,0}^{(0)\alpha}(y)\,, {\cal J}_g^{(0)}(\bn\mcdot p_2, \bn\mcdot p_1, x)\,,
\bar\chi_{n,p}^{(0)\beta}(0))|0\rangle \\
& & = i\int \mbox{d}^4 x
\langle 0|
\mbox{T}(\Phi_{n,0}^{(0)\alpha}(y)\,, {\cal J}_g^{(0)}(\bn\mcdot p_1 - \bn\mcdot q,
\bn\mcdot p_1, x)\,,
\bar\chi_{n,p}^{(0)\beta}(0))|0\rangle\nonumber\\
& & \equiv
i\delta_{\bn\mcdot p, \bn\mcdot q}\delta_{\bn\mcdot p, \bn\mcdot p_1}
\int\frac{d^4 k}{(2\pi)^4}
e^{-ik\mcdot y} J_\ell (\bn\mcdot p; n\mcdot k)
\left[\frac{\bnslash\nslash}{4}\epsslash_\perp^*\right]^{\alpha\beta}\,.\nonumber
\eea
This defines a new collinear function $J_\ell(\bn\mcdot p; n\mcdot k)$, similar
to $J(\bn\mcdot p; n\mcdot k)$ introduced in (\ref{Jdef}).
The operator ${\cal J}_g^{(0)}$ is related to ${\cal J}_g$ introduced in (\ref{J0})
by the field redefinition
(\ref{redef}) and is given explicitly by
%
\bea
{\cal J}_g^{(0)}(\bn\mcdot p_2, \bn\mcdot p_1 ; y) = 
\bar\chi^{(0)}_{n,p_2} \epsslash^*_{\perp}\frac{1}{\bn\mcdot \cal P} 
\Phi^{(0)}_{n,p_1} + \bar\Phi_{n,p_2}^{(0)}
\frac{1}{\bn\mcdot {\cal P} -\bn\mcdot q} \epsslash^* _{\perp}\chi^{(0)}_{n,p_1}\,.
\eea
In contrast to the function $J$ which starts at tree level, $J_\ell$
starts at $O(\alpha_s)$ since the sub-leading operator ${\cal
L}_{q\xi}^{(1)}$ has only Feynman rules with at least one collinear
gluon. The three graphs contributing to $J_\ell$ at lowest order are
shown in Figs.~\ref{collTprod}(e)-(g). Their computation is described in the
Appendix, and the result reads (in Feynman gauge)
\bea
J_\ell(\bn\mcdot p, n\mcdot k) = \frac{1}{n\mcdot k}\frac{\alpha_s C_F}{4\pi}
\left(- L^2 +3 L -8 + \frac{\pi^2}{6}\right)\,.
\label{jetfunctionJl}
\eea
Note that, using the same argument as in the $J(\bn\mcdot p; n\mcdot k)$
case, the collinear function $J_\ell(\bn\mcdot p; n\mcdot k)$ also depends 
on $k$ only through $n\mcdot k$ at all orders in perturbation theory.

Inserting Eq.~(\ref{Jell}) into the factorized matrix element
(\ref{Aeff}), multiplying with the ultra-soft matrix element on the $B$
meson, and performing the integration over $x$ gives $A_{\rm
eff}^{i(2)}$ similar to $A_{\rm eff}^{i(1)}$ but with the replacement
$J(2E_\gamma, n\cdot k) \to J_\ell(2E_\gamma, n\cdot k)$. This
determines $A_{\rm eff}^{i}= A_{\rm eff}^{i(1)}+A_{\rm eff}^{i(2)}$
to be given by an expression similar to (\ref{Aeff4}) with the
replacement $J\to J+J_\ell$.

The final result for the $\bar B\to \gamma e\bar\nu$ decay amplitude
is obtained by inserting the factorized soft-collinear matrix elements
as in Eq.~(\ref{Aeff4}) into the expansion (\ref{hard}) and performing
the sum over all Dirac structures for the current in the effective
theory. One defines separate form factors $f_{V,A}(E_\gamma)$ for the
vector and axial currents in QCD by (with the convention $\varepsilon^{0123}=1$)
\bea
\frac{1}{e}\langle \gamma(q,\varepsilon) |\bar q\gamma_\mu b|\bar B(v)\rangle &=&
i\epsilon_{\mu\alpha\beta\delta} \varepsilon_\alpha^* v_\beta q_\delta f_V(E_\gamma)\\
\frac{1}{e}\langle \gamma(q,\varepsilon) |\bar q\gamma_\mu\gamma_5 b|\bar B(v)\rangle 
&=&
[q_\mu (v\mcdot \varepsilon^*) - \varepsilon_\mu^* (v\mcdot q)] f_A(E\gamma)\\
& &+ (v\mcdot \varepsilon^*) v_\mu \frac{1}{v\mcdot q} f_B m_B\nonumber\,.
\eea
The second term in the matrix element of the axial current is required by
gauge invariance \cite{gauge} and cancels the photon emission amplitude from
the charged lepton line.
With these definitions, one finds the following explicit results for the
$B\to \gamma$ form factors at leading order in $1/E_\gamma$
\bea\label{fVAfinal}
f_V(E_\gamma) &=& C^{(v)}_1(2E_\gamma, \mu)
\frac{Q_q}{E_\gamma} I(\mu)\; , \\
\label{fVAfinal1}
f_A(E_\gamma) &=&  C^{(a)}_1(2E_\gamma, \mu)
\frac{Q_q}{E_\gamma} I(\mu) \; ,
\eea
where the usoft-collinear part of the matrix element is contained in the
integral
\bea
\label{integ}
I(\mu) &\equiv& \int \frac{dn\mcdot k}{4\pi} 
(J(2E_\gamma; n\mcdot k) + J_\ell(2E_\gamma, n\mcdot k)) \; \psi_+(n\mcdot k).
\eea

The results
(\ref{fVAfinal})-(\ref{integ}) can be summarized at one-loop by
writing
\bea
\label{1loopFF}
f_V(E_\gamma) = f_A(E_\gamma) = \frac{Q_q}{E_\gamma}\int \frac{dn\mcdot k}{4\pi}
\frac{1}{n\mcdot k}
\left\{ 1 + \frac{\alpha_s C_F}{4\pi}T^{(1)}(n\mcdot k,\mu)\right\} \psi_+(n\mcdot k,\mu)
\eea
where the one-loop correction to the hard scattering kernel
$T^{(1)}(n\mcdot k,\mu)$ is given by
\bea
\label{1loopT}
T^{(1)} (n\mcdot k, \mu) &=& -\frac12 \log^2\left(\frac{\mu^2}{m_b^2}\right) + 
\log\left(\frac{\mu^2}{m_b^2}\right)
\left(2\log\frac{2E_\gamma}{m_b} - \frac52\right)
- 2\frac{3E_\gamma - m_b}{m_b - 2E_\gamma}\log\frac{2E_\gamma}{m_b}\\
& & + \log^2\left(\frac{2E_\gamma n\mcdot k}{\mu^2}\right) 
- 2\log^2\left(\frac{2E_\gamma}{m_b}\right) - 
2\mbox{Li}_2\left(1 - \frac{2E_\gamma}{m_b}\right)
- \frac{\pi^2}{4} - 7\,.\nonumber
\eea
This expression coincides with the one-loop results presented in Ref.~\cite{DGS}.

Similar results are obtained for the form factors of the tensor current,
which can be defined as 
\bea
\frac{1}{e}
\langle \gamma(q,\varepsilon)|\bar q\sigma_{\mu\nu} b|\bar B(p)\rangle &=&
g_+(E_\gamma) i\varepsilon_{\mu\nu\alpha\beta} \varepsilon^{*\alpha} (p+q)^\beta +
g_-(E_\gamma) i\varepsilon_{\mu\nu\alpha\beta} \varepsilon^{*\alpha} (p-q)^\beta\\
&-& 2h(E_\gamma) (\varepsilon^*\mcdot p) 
i\varepsilon_{\mu\nu\alpha\beta} p^\alpha q^\beta
\nonumber
\eea
The explicit leading order results for these form factors are
\bea\label{t1}
g_+(E_\gamma) &=& \frac12 \left\{C_1^{(t)}(2E_\gamma,\mu) +
C_2^{(t)}(2E_\gamma,\mu) + \left(1-\frac{E_\gamma}{m_B}\right) C_3^{(t)}(2E_\gamma,\mu)
\right\} \frac{Q_q}{E_\gamma} I(\mu)\\
\label{t2}
g_-(E_\gamma) &=& -\frac12 \left\{C_1^{(t)}(2E_\gamma,\mu) +
C_2^{(t)}(2E_\gamma,\mu) + \left(1+\frac{E_\gamma}{m_B}\right) C_3^{(t)}(2E_\gamma,\mu)
\right\} \frac{Q_q}{E_\gamma} I(\mu)\\
\label{t3}
h(E_\gamma) &=& \frac{1}{2m_B^2} C_3^{(t)}(2E_\gamma,\mu) \frac{Q_q}{E_\gamma} I(\mu)
\eea

Writing the form factors in this form makes explicit their
factorization into a hard ($C_i$), collinear ($J$) and ultra-soft
($\psi_+$) components.  This result gives a field-theoretical
interpretation of the contributions coming from different momentum
regions. Corrections to factorization can be expected from
subleading operators in the matching of the currents and Lagrangian insertions,
for which the field redefinition (\ref{redef}) fails to decouple ultrasoft 
from collinear fields \cite{bpssoft}.

In the numerical evaluation of the form factors, a convenient choice
for the scale $\mu$ is $\mu^2 \sim 2E_\gamma \Lambda$, for which the
logs appearing in the collinear functions $J, J_\ell$ are
minimized. With this choice, these logs are shifted instead into the
Wilson coefficients $C_i(2E_\gamma)$, where they can be resummed using
renormalization group (RG) methods. It has been shown in
\cite{bfl,bfps} that the Wilson coefficients $C_i(\bn\mcdot p,\mu)$ satisfy RG
equations of the form
\bea
\mu\frac{d}{d\mu}C_i(\bn\mcdot p, \mu) = \gamma(\bn\mcdot p/\mu) C_i(\bn\mcdot p, \mu)
\eea
with a universal anomalous dimension $\gamma$. The form of the
resummed result has been discussed in detail in \cite{bfps} to which
we refer for explicit results.
A similar resummation can be performed using
the methods of \cite{GPK,KS}, which are equivalent to the effective theory
approach. At leading order, the resummation of the Sudakov double logs 
$\log^2(2E_\gamma/\mu)$
has been given in \cite{KPY} using the approach of \cite{GPK,KS}, 
and a next-to-leading analysis was done recently in \cite{DGS} using SCET methods.

The factorized form (\ref{fVAfinal})-(\ref{fVAfinal1}) exhibits a
symmetry relation between radiative leptonic form factors of different
currents. At $O(\lambda^0)$, the only difference between
$f_V(E_\gamma)$ and $f_A(E_\gamma)$ comes from the Wilson coefficients
of the heavy light currents. However, it has been shown in \cite{bfps}
that, for massless collinear quarks, the Wilson coefficients $C^{(v)}_1$ and
$C^{(a)}_1$ are equal to all orders in $\alpha_s$ as a consequence of
the chiral symmetry of the effective theory. This proves the symmetry
relation $f_V(E_\gamma) = f_A(E_\gamma)$ to all orders in $\alpha_s$,
confirming a result conjectured in \cite{KPY} from an explicit
one-loop computation. Analogous relations can be given for the form
factors of the tensor current following from (\ref{t1})-(\ref{t3}).

Finally, we comment on the relation of our results to a factorization
formula for $B\to \gamma e\nu$ conjectured in \cite{KPY} from a
one-loop computation. The main difference is the simpler one-dimensional form
of the convolution formula for the form factors (\ref{fVAfinal}),
(\ref{integ}) (and also in \cite{DGS}), compared to the three-dimensional 
factorization formula
proposed in \cite{KPY}. This is due to a different definition for the
B meson wave function adopted in \cite{KPY}, which shifts part of the
usoft matrix element ($k_\perp$-dependent) into the hard scattering amplitude $T$.

Another difference which must be taken into account when comparing
with the results of \cite{KPY} is the hierarchy $m_b \gg 2E_\gamma \gg
\Lambda$ adopted in that paper. This corresponds to expanding the hard
scattering kernel $T$ in $2E_\gamma/m_b$ and keeping only the leading
term.  On the other hand, the more general approach used here allows
$m_b, 2E_\gamma$ to be of comparable magnitude, subject only to the
condition $m_b, 2E_\gamma \gg \Lambda$.

\section{Conclusions}

We have studied in this paper the consequences of the presence of
two considerably different scales, $2E_\gamma \sim m_b$ and $\Lambda_{QCD}$.
Although this is a general feature of 
radiative $B$ decays involving an energetic photon, the basic physical features 
of such a problem can be seen on the simpler case of the leptonic radiative
decay $B\to \gamma e\nu$. The presence of a small parameter
$\lambda$, where $\lambda^2 = \Lambda_{QCD}/m_b$, suggests
using the soft-collinear effective theory which is tailored to
systematically treat the expansion in $\lambda$.

We discussed in this paper the most general form of the SCET operators 
contributing to the leptonic radiative decay $\bar B\to\gamma e\bar\nu$ at 
leading order in $\lambda$. 
Using this result, we showed that the form factors describing this
decay factorize into hard, collinear and soft factors to all orders in 
perturbation theory. 
Further work is needed to complete the proof of the factorization
relation for a generic radiative B decay. One can hope that the application of the
soft-collinear effective theory along the lines of the present paper can be 
fruitful in unravelling the general form of such a result.

\section{Acknowledgments}
We are grateful to Grisha Korchemsky and Iain W.~Stewart for useful discussions. 
We thank Andri Hardmeier for many discussions and suggestions.
E.L. acknowledges financial support from the Alexander Von Humboldt
Foundation. The research of D.~P. has been supported by the DOE under Grant
No. DOE-FG03-97ER40546 and by the 
U.S. National Science Foundation Grant PHY-9970781.
E.~L. and D.~W. are supported by the Swiss National Science Foundation.

\appendix
\section{One-loop matching of the ultra-soft--collinear current}
\begin{figure}[b!]
\begin{center}
  \includegraphics[width=2.0in]{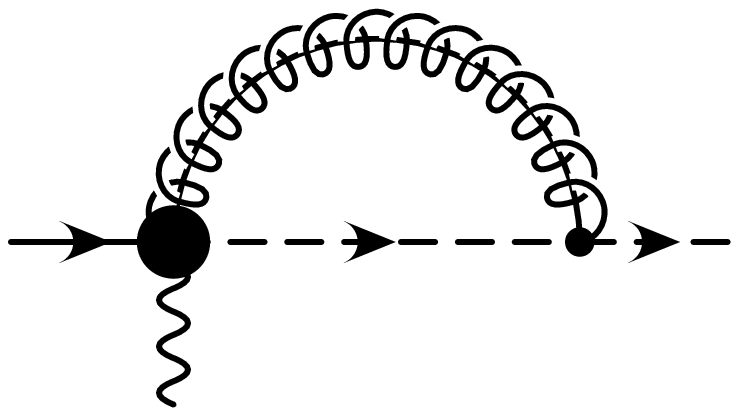}
 \includegraphics[width=2.0in]{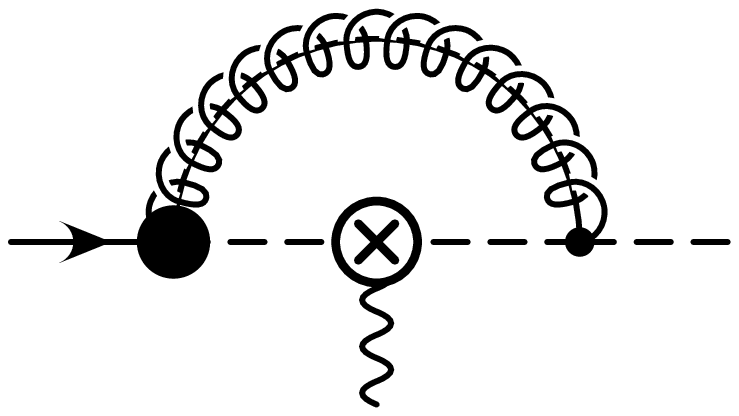}
 \includegraphics[width=2.0in]{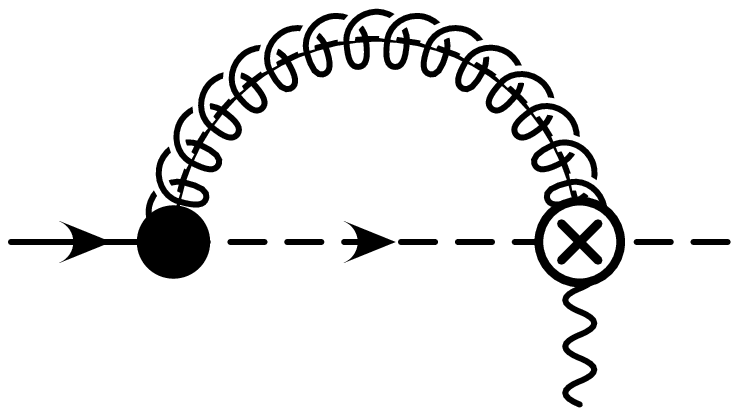}\\
(a) \hspace{1.8in} (b) \hspace{1.8in} (c)
\end{center}
{\caption{Collinear graphs for photon emission from a light quark
in the effective theory. The wiggly line denotes the photon.
The graph (a) corresponds to the local operator in the matching,
and the remaining two (b), (c) come from the nonlocal term.
The blob in (a) denotes the $\bar\xi \epsslash_\perp^* Wq$ operator;
in (b), (c) the shaded blob denotes the ${\cal L}^{(1)}_{\xi q}$ 
subleading Lagrangian
and the circled cross is the insertion of ${\cal J}_g$.
\label{collfig}}}
\end{figure}

We consider here in some detail the coupling of an energetic
photon $\gamma(q,\varepsilon)$ to an ultra-soft quark $q$, turning it
into a collinear quark $\xi_n$. Proceeding in analogy to the
heavy-light current \cite{bfl,bfps} one would be led to match this
coupling onto a local operator in the effective theory. At order
$O(\lambda^0)$ the most general operator compatible with collinear
gauge invariance is $\bar \xi_n W\epsslash^* q$. In Sec.~III it
was shown that the complete matching can include also a nonlocal
term and it reads
\bea\label{AA1}
\bar q\epsslash^*_\perp q \to
\bar \xi_n \epsslash^*_\perp W q + iT\{ {\cal J}_g\,, {\cal L}^{(1)}_{\xi q}\}\,.
\eea
The operators appearing in the nonlocal term are defined as
\bea
{\cal J}_g &=& \bar\xi_{n,p-q} \left[
\epsslash_\perp^* \frac{1}{\bn\mcdot {\cal P} + g\bn\mcdot A_n }
(\cPslash_\perp + g\Aslash_\perp) +
(\cPslash_\perp + g\Aslash_\perp)
\frac{1}{\bn\mcdot {\cal P} -\bn\mcdot q + g\bn\mcdot A_n}\epsslash_\perp^*\right]
\frac{\bnslash}{2}\xi_{n,p}\nn\\
{\cal L}^{(1)}_{\xi q} &=& \bar\xi_n (\cPslash_\perp + g\Aslash_\perp)
Wq\,.
\label{AA2}
\eea
The SCET diagrams involving collinear gluons that describe the one-loop
matching of the ultra-soft-collinear current are given in Fig.~\ref{collfig};
we do not draw the diagram with one ultra-soft gluon since it has the same
topology as in full QCD. The graph in Fig.~\ref{collfig}(a) is the
contribution analogous to the one present in the heavy-light
correction; it stems from the first term in Eq.~(\ref{AA1}) and
the two vertices denote, respectively, an insertion of the
$\bar\xi_n \epsslash_\perp^* Wq$ operator (the blob) and of the leading
order Lagrangian ${\cal L}^{(0)}$. The two remaining graphs,
Figs.~\ref{collfig}(b)-(c), appear only for the light-light case and contain one
insertion of the sub-leading Lagrangian ${\cal L}^{(1)}_{\xi q}$ and one
insertion of the operator ${\cal J}_g$ (the diagram in Fig.~\ref{collfig}(b)
contains also an insertion of the leading order lagrangian).
Using Eqs.~(\ref{AA1}) and (\ref{AA2}) it is immediate to extract the Feynman
rules for the various vertices that appear in Fig.~\ref{collfig}.

In the explicit computation of the diagrams we
assume that the photon moves along the $n$ direction with
light cone momentum components $q = (n\mcdot q,\bn\mcdot q,q_\perp) =
(0,\bn\mcdot q, 0)$. The momenta of the ingoing ultrasoft quark and of the
outgoing collinear quark are $p$ and $p'=p-q$, respectively. The ultrasoft
quark is taken on-shell with $p^2=m^2$.

The contribution of the local term in (\ref{AA1}), Fig.~\ref{collfig}(a), is identical to the
one-loop correction to the heavy-light vertex~\cite{bfl}, and its computation
gives (we use everywhere in the following the Feynman gauge)
\bea
\Gamma_{\mbox{\tiny C}}^{(a)}
&=&
-2 i g_s^2 C_F \mu^{2\epsilon} 
\int \frac{\mbox{d}^d k}{(2\pi)^d} \frac{\bar n \mcdot (k-q)}{(p-q+k)^2 \;
\bar n \mcdot k \; k^2}[\bar \xi_n(p')\epsslash^*_\perp u(p)] \nn \\
& =&
-\frac{2\alpha_s C_F}{4\pi}
\frac{\Gamma(\epsilon)\Gamma(-\epsilon)\Gamma(2-\epsilon)}{\Gamma(2-2\epsilon)}
\left(\frac{n\mcdot p \bn\mcdot q}{\mu^2}\right)^{-\epsilon}
[\bar \xi_n(p')\epsslash^*_\perp u(p)] \nn \\
&=&
\frac{\alpha_s C_F}{4\pi} \left(  \frac{2}{\epsilon^2} -\frac{2L -2 }{\epsilon}
+L^2 - 2 L - \frac{\pi^2}{6}  + 4
\right)[\bar \xi_n(p')\epsslash^*_\perp u(p)]
\eea
where $d=4 -2 \epsilon$, $L =\log (n\mcdot p \bn\mcdot q/\mu^2_E) = \log (-{p'}^2/\mu^2_E)$
and $\mu^2_E = \mu^2 e^{\gamma_E - \log 4 \pi}$. The result contains a double
logarithm \cite{bfl}. In an analogous way we compute the contributions from the nonlocal term in
(\ref{AA1}), Figs.~\ref{collfig}(b)-(c), and obtain
\bea
\Gamma_{\mbox{\tiny C}}^{(b+c)} =
\frac{\alpha_s C_F}{4\pi} \left(  -\frac{2}{\epsilon^2} - \frac{3-2L}{\epsilon}
- L^2 + 3L + \frac{\pi^2}{6} - 8
\right) [\bar \xi_n(p')\epsslash^* u(p)]\; .
\eea
Finally, the sum of the three diagrams reads
\bea
\Gamma_{\mbox{\tiny C}} =
\frac{\alpha_s C_F}{4\pi} \left( -\frac{1}{\epsilon} + \log\left( \frac{-{p'}^2}{\mu^2_E}\right) - 4
\right) [\bar \xi_n(p')\epsslash^*_\perp u(p)]
\label{Gcoll}
\eea
where we restored the logarithm. Note that in Eq.~(\ref{Gcoll}) the double logarithms
cancel.

To complete the computation of the vertex correction in SCET we need to include
also the usoft diagram that comes at one-loop order only
from one insertion of the local term in (\ref{AA1}), and is equal to
\bea
\Gamma_{\mbox{\tiny S}}
&=&
-2ig^2 C_F \mu^{2\epsilon}\int\frac{\mbox{d}^d k}{(2\pi)^d}
\frac{1}{[(k+p)^2-m^2 +i\varepsilon](k^2 + i\varepsilon)}
[\bar \xi(p')\epsslash^*_\perp u(p)] \nn \\
&=&
\frac{2\alpha_s C_F}{4\pi}\frac{\Gamma(\epsilon)}{1-2\epsilon}
\left(\frac{m^2}{\mu^2}\right)^{-\epsilon}[\bar \xi(p')\epsslash^*_\perp u(p)] \nn\\
&=&
 \frac{\alpha_s C_F}{4\pi}\left(\frac{2}{\epsilon} - 2\log\left(\frac{m^2}{\mu_E^2}\right)
+ 4\right) [\bar \xi(p')\epsslash^*_\perp u(p)]
\label{Gsoft}
\eea
where $m$ is the mass of the ultrasoft quark that we use as IR regulator.

The total SCET contributions to the usoft-collinear vertex is
\bea
\Gamma_{\mbox{\tiny SCET}} =
 \frac{\alpha_s C_F}{4\pi}\left[ \frac{1}{\epsilon}
+ \log \left( \frac{-{p'}^2}{\mu^2_E}\right)- 2\log\left(\frac{m^2}{\mu_E^2}\right)
\right] [\bar \xi(p')\epsslash^*_\perp u(p)]
\label{GSCET}
\eea

Finally, computing the vertex correction in full QCD, we obtain
\bea
\Gamma_{\mbox{\tiny QCD}}
&=&
-i g_s^2 C_F \mu^{2\epsilon} \int \frac{\mbox{d}^d k}{(2\pi)^d}
\frac{
\gamma^\mu (\ppslash + \kslash +m) \epsslash^* (\pslash +\kslash+m) \gamma_\mu
}
{
((p+k)^2-m^2) \; k^2 \; ((p'+k)^2-m^2)
} \nn \\
&=&
\Gamma_{\mbox{\tiny SCET}} + 
O\left(\frac{m^2}{(\bn\mcdot q)^2}, \frac{{p'}^2}{(\bn\mcdot q)^2}\right) \; .
\eea
This exercise shows that the IR behavior of QCD is exactly
reproduced in the SCET at the one-loop order; moreover, it also
shows to what extent the effective theory computation reproduces
the full QCD result.

Before concluding this appendix let us describe briefly the
computation of the jet functions $J$ and $J_l$ introduced in
Eqs.~(\ref{jetfunctionJ}) and (\ref{jetfunctionJl}).
The diagrams that produce the latter are exactly those drawn in
Figs.~\ref{collfig}(b)-(c) in which the ultra-soft field on the
left is dropped and a collinear propagator is inserted on the
right (as can be readily seen expanding explicitly the T-product
in Eq.~(\ref{Jell})). In fact, after subtracting the divergences
(remember that $J_l$ is defined as the vacuum-to-vacuum matrix
element of a T-product of effective theory fields)
$\Gamma_{\mbox{\tiny C}}^{(b+c)}$ exactly reproduces
Eq.~(\ref{jetfunctionJl}).

The leading order contribution to $J$ is simply the propagator of
the collinear field while the $O(\alpha_s)$ correction is given by
the sum of three graphs. The first diagram is given by the diagram in
Fig.~\ref{collfig}(a) in which the ultra-soft field on the left is
dropped and a collinear propagator is inserted on the right.
The second diagram is obtained from the first one by moving the
collinear propagator to the left side. These two diagrams yield
identical results and their contribution to $J$ is
$1/(n\mcdot k) \alpha_s C_F /(4\pi) (2 L^2 -4L+8 -\pi^2/3)$.
The third diagram is simply the one loop correction to the quark
propagator (analogous to QCD). Its contribution was computed in
Eq.~(36) of Ref.~\cite{bfps} and yields
$1/(n\mcdot k) \alpha_s C_F /(4\pi) (L-1)$.
The sum of the three graphs reproduces Eq.~(\ref{jetfunctionJ}).


\end{document}